\documentclass[twocolumn,english,prl, nofootinbib]{revtex4-1}
\usepackage[T1]{fontenc}
\usepackage[latin9]{inputenc}
\usepackage{color}
\usepackage{babel}
\usepackage{units}
\usepackage{amsmath}
\usepackage{amssymb}
\usepackage{graphicx}
\usepackage[unicode=true,pdfusetitle,
 bookmarks=true,bookmarksnumbered=false,bookmarksopen=false,
 breaklinks=false,pdfborder={0 0 0},pdfborderstyle={},backref=false,colorlinks=true]
 {hyperref}
\hypersetup{
 urlcolor=orange,linkcolor=blue,citecolor=red}

\makeatletter
\usepackage{tikz}
\usepackage{color}
\usepackage{newtxtext}
\usepackage{newtxmath}
\usepackage{dsfont}

\makeatother

\begin{document}
\title{Properties of the geometry of solutions and capacity of multi-layer
neural networks with Rectified Linear Units activations}
\author{Carlo Baldassi}
\affiliation{Artificial Intelligence Lab, Institute for Data Science and Analytics,
Bocconi University, Milano, Italy}
\author{Enrico M. Malatesta}
\email{enrico.malatesta@unibocconi.it}

\affiliation{Artificial Intelligence Lab, Institute for Data Science and Analytics,
Bocconi University, Milano, Italy}
\author{Riccardo Zecchina}
\affiliation{Artificial Intelligence Lab, Institute for Data Science and Analytics,
Bocconi University, Milano, Italy}
\begin{abstract}
Rectified Linear Units (ReLU) have become the main model for the neural
units in current deep learning systems. This choice has been originally
suggested as a way to compensate for the so called vanishing gradient
problem which can undercut stochastic gradient descent (SGD) learning
in networks composed of multiple layers. Here we provide analytical
results on the effects of ReLUs on the capacity and on the geometrical
landscape of the solution space in two-layer neural networks with
either binary or real-valued weights. We study the problem of storing
an extensive number of random patterns and find that, quite unexpectedly,
the capacity of the network remains finite as the number of neurons
in the hidden layer increases, at odds with the case of threshold
units in which the capacity diverges. Possibly more important, a large
deviation approach allows us to find that the geometrical landscape
of the solution space has a peculiar structure: While the majority
of solutions are close in distance but still isolated, there exist
rare regions of solutions which are much more dense than the similar
ones in the case of threshold units. These solutions are robust to
perturbations of the weights and can tolerate large perturbations
of the inputs. The analytical results are corroborated by numerical
findings.
\end{abstract}
\maketitle
Artificial Neural Networks (ANN) have been studied since decades and
yet only recently they have started to reveal their potentialities
in performing different types of massive learning tasks~\citep{lecun2015deep}.
Their current denomination is Deep Neural Networks (DNN) in reference
to the the choice of the architectures, which typically involve multiple
interconnected layers of neuronal units. Learning in ANN is in principle
a very difficult optimization problem, in which ``good'' minima
of the learning loss function in the high dimensional space of the
connection weights need to be found. Luckily enough, DNN models have
evolved rapidly, overcoming some of the computational barriers that
for many years have limited their efficiency. Important components
of this evolution have been the availability of computational power
and the stockpiling of extremely rich data sets.

The features on which the various modeling strategies have intersected,
besides the architectures, are the choice of the the loss functions,
the transfer functions for the neural units and the regularization
techniques. These improvements have been found to help the convergence
of the learning processes, typically based on Stochastic Gradient
Descent (SGD)~\citep{bottou2010large}, and to lead to solutions
which can often avoid overfitting even in over parametrized regimes.
All these results pose basic conceptual questions which need to find
a clear explanation in term of the optimization landscape.

Here we study the effects that the choice of the Rectified Linear
Units (ReLU) for the neurons \citep{hahnloser2000digital} has on
the geometrical structure of the learning landscape. ReLU is one of
the most popular non-linear activation functions and it has been extensively
used to train DNN, since it is known to dramatically reduce the training
time for a typical algorithm~\citep{glorot2011deep}. It is also
known that another major benefit of using ReLU is that it does not
produce the vanishing gradient problem as other common transfer functions
(e.g. the $\mathrm{tanh}$) \citep{hochreiter1991untersuchungen,glorot2011deep}.
We study ANN models with one hidden layer storing random patterns,
for which we derive analytical results that are corroborated by numerical
findings. At variance with what happens in the case of threshold units,
we find that models built on ReLU functions present a critical capacity,
i.e. the maximum number of patterns per weight which can be learned,
that does not diverge as the number of neurons in the hidden unit
is increased. At the same time we find that below the critical capacity
they also present wider dense regions of solutions. These regions
are defined in terms of the volume of the weights around a minimizer
which do not lead to an increase of the loss value (e.g. number of
errors) \citep{baldassi_unreasonable_2016}. For discrete weights
this notions reduces to the so called Local Entropy \citep{baldassi2015subdominant}
of a minimizer. We also check analytically and numerically the improvement
in the robustness of these solutions with respect to both weight and
input perturbations. 

Together with the recent results on the existence of such wide flat
minima and on the effect of choosing particular loss functions to
drive the learning processes toward them \citep{baldassi2019shaping},
our result contributes to create a unified framework for the learning
theory in DNN, which relies on the large deviations geometrical features
of the accessible solutions in the overparametrized regime.

\noindent \textit{The model}. We will consider a two-layer neural
network with $N$ input units, $K$ neurons in the hidden layer and
one output. The mapping from the input to the hidden layer is realized
by $K$ non-overlapping perceptrons each having $N/K$ weights. Given
$p=\alpha N$ inputs $\xi^{\mu}$ labeled by index $\mu=\left\{ 1,\dots,p\right\} $,
the output of the network for each input $\mu$ is computed as 
\begin{equation}
\sigma_{\text{out}}^{\mu}=\mathrm{sgn}\left(\frac{1}{\sqrt{K}}\sum_{l=1}^{K}c_{l}\,\tau_{l}^{\mu}\right)=\mathrm{sgn}\left(\frac{1}{\sqrt{K}}\sum_{l=1}^{K}c_{l}\,g\left(\lambda_{l}^{\mu}\right)\right)\,,\label{eq:out}
\end{equation}
where $\lambda_{l}^{\mu}$ is the input of the $l$ hidden unit i.e.
$\lambda_{l}^{\mu}=\sqrt{\frac{K}{N}}\sum_{i=1}^{N/K}W_{li}\xi_{li}^{\mu}$
and $W_{li}$ is the weight connecting the input unit $i$ to the
hidden unit $l$. $c_{l}$ is the weight connecting hidden unit $l$
with the output; $g$ is a generic activation function. In the following
we will mainly consider two particular choices of activation functions
and of the weights $c_{l}$. In the first one we take the sign activation
$g\left(\lambda\right)=\mathrm{sgn}\left(\lambda\right)$ and we fix
to 1 all the weights $c_{l}$ (in general their sign can be absorbed
into the weights $W_{li}$). The $K=1$ version of this model is the
well-known perceptron and it has been extensively studied since the
'80s by means of the replica and cavity methods~\citep{gardner1988The,gardner1988optimal,mezard1989space}
used in spin glass theory~\citep{BOOKNishimori2001}. The $K>1$
case is known as the tree-committee machine that has been studied
in the '90s~\citep{barkai1992broken,engel1992storage}. In the second
model we will use the ReLU activation function that is defined with
$g\left(\lambda\right)=\max\left(0,\lambda\right)$, and since the
output of this transfer function is always non-negative we will fix
half of the weights $c_{l}$ to +1 and the remaining half to $-1$.
Given a training set composed by random i.i.d patterns $\xi^{\mu}\in\left\{ -1,1\right\} ^{N}$
and labels $\sigma^{\mu}\in\left\{ -1,1\right\} $ and defining $\mathbb{X}_{\xi,\sigma}\left(W\right)\equiv\prod_{\mu}\theta\left(\frac{\sigma^{\mu}}{\sqrt{K}}\sum_{l=1}^{K}c_{l}\,\tau_{l}^{\mu}\right)$,
the weights that correctly classify the patterns are those for which
$\mathbb{X}_{\xi,\sigma}\left(W\right)=1$. Their volume (or number)
is therefore~\citep{gardner1988The,gardner1988optimal} 
\begin{equation}
Z=\int\!\!\mathrm{d}\mu\left(W\right)\,\mathbb{X}_{\xi,\sigma}\left(W\right)\,,\label{GardnerVolume}
\end{equation}
where $\mathrm{d}\mu(W)$ is a measure over the weights $W$. In this
study two constraints over the weights will be considered. The spherical
constraint where for every $l\in\left\{ 1,\dots,K\right\} $, we have
$\sum_{i}W_{li}^{2}=\nicefrac{N}{K}$, i.e. every sub-perceptron has
weights that live on the hypersphere of radius $\sqrt{\nicefrac{N}{K}}$.
The second constraint we will use is the binary one, where for every
$l\in\left\{ 1,\dots,K\right\} $ and $i\in\left\{ 1,\dots,\nicefrac{N}{K}\right\} $
we have $W_{li}\in\left\{ -1,1\right\} $. We are interested in the
large $K$ limit for which we will be able to compute analytically
the capacity of the model for different choices of transfer function,
to study the typical distances between absolute minima and to perform
the large deviation study giving the local volumes associated to the
wider flat minima.

\noindent \textit{Critical Capacity}. We will analyze the problem
in the limit of a large number $N$ of input units. The standard scenario
in this limit is that there is a sharp threshold $\alpha_{c}$ such
that for $\alpha<\alpha_{c}$ the probability of finding a solution
is 1 while for $\alpha>\alpha_{c}$ the volume of synapses is empty.
$\alpha_{c}$ is therefore called \emph{critical capacity} since it
is the maximum number of patterns per synapses that one can store
in a neural network. The critical capacity of the mode, for a generic
choice of the transfer function, can be evaluated computing the free
entropy $\mathcal{F}\equiv\frac{1}{N}\left\langle \ln Z\right\rangle _{\xi,\sigma}$,
where $\left\langle \cdot\right\rangle _{\xi,\sigma}$ denotes the
average over the patterns, using the replica method; one finds 
\begin{equation}
\mathcal{F}=\mathcal{G}_{S}+\alpha\,\mathcal{G}_{E}\,.
\end{equation}
$\mathcal{G}_{S}$ is the entropic term, which represents the logarithm
of the volume at $\alpha=0$, where there are no constraints induced
by the training set; this quantity is independent on $K$ and it is
affected only by the binary or spherical nature of the weights. $\mathcal{G}_{E}$
is the energetic term and it represents the logarithm of the fraction
of solutions. Moreover it depends on the order parameters $q_{l}^{ab}\equiv\frac{K}{N}\sum_{i}W_{li}^{a}W_{li}^{b}$
which represent the overlap between sub-perceptrons $l$ of two different
replicas $a$ and $b$ of the machine. Using a replica-symmetric (RS)
ansatz, in which we assume $q_{l}^{ab}=q$ for all $a,b,l$, and in
the large $K$ limit, $\mathcal{G}_{E}$ is 
\begin{equation}
\mathcal{G}_{E}=\int\!\!Dz_{0}\,\ln H\left(-\sqrt{\frac{\Delta-\Delta_{-1}}{\Delta_{2}-\Delta}}z_{0}\right)\label{G_E_RS}
\end{equation}
where $Dz\equiv\frac{\mathrm{d}z}{\sqrt{2\pi}}e^{-z^{2}/2}$ and $H\left(x\right)\equiv\int_{x}^{\infty}Dz$.
This expression is equivalent to that of the perceptron (i.e. $K=1$),
the only difference being that the order parameters are replaced by
effective ones that depend on the general activation function used
in the machine~\citep{barkai1992broken}. In~(\ref{G_E_RS}) we
have called these effective order parameters as $\Delta_{-1}$, $\Delta$
and $\Delta_{2}$, see the appendix~\ref{subsec:RS} for details,
and in the perceptron they are 0, $q$ and $1$ respectively.

In the binary case the critical capacity is always smaller than 1
and it is identified with the point where the RS free entropy $\mathcal{F}$
vanishes. This condition requires 
\begin{equation}
\alpha_{c}=\frac{\frac{\hat{q}}{2}(1-q)-\int Du\ln\left(2\cosh\left(\sqrt{\hat{q}}u\right)\right)}{\int Dz_{0}\,\ln H\left(-\sqrt{\frac{\Delta-\Delta_{-1}}{\Delta_{2}-\Delta}}z_{0}\right)}
\end{equation}
where $\hat{q}$ is the conjugated parameter of $q$. $q$ and $\hat{q}$
are found by solving their associated saddle point equations (details
in appendix~\ref{subsec:RS}). Solving these equations one finds
for the ReLU case $\alpha_{c}=0.9039(9)$ which is a smaller value
than in the sign activation function case, where one gets $\alpha_{c}=0.9469(5)$
as shown in~\citep{barkai1992broken}.

In the spherical case the situation is different since the capacity
is not bounded from above. Previous works~\citep{engel1992storage,barkai1992broken}
have shown, in the case of the sign activations, that the RS estimate
of the critical capacity diverges with the number of neurons in the
hidden layer as $\alpha_{c}\simeq\left(\frac{72K}{\pi}\right)^{1/2}$,
violating the Mitchison-Durbin bound~\citep{mitchison1989bounds}.
The reason of this discrepancy is due to the fact that the Gardner
volume disconnects before $\alpha_{c}$ and therefore replica-symmetry
breaking (RSB) takes place. Indeed, the instability of the RS solution
occurs at a finite value $\alpha_{\text{AT}}\simeq2.988$ at large
$K$. A subsequent work~\citep{monasson1995weight} based on multifractal
techniques derived the correct scaling of the capacity with $K$ as
$\alpha_{c}\simeq\frac{16}{\pi}\sqrt{\ln K}$, which saturates the
Mitchison-Durbin bound.

In the case of the ReLU functions the RS estimate of the critical
capacity is obtained simply performing the $q\to1$ limit, as for
the perceptron. Quite surprisingly, if the activation function is
such that $\Delta_{2}-\Delta\simeq\delta\Delta\left(1-q\right)$ for
$q\to1$, with $\delta\Delta$ a finite proportionality term, the
RS estimate of the critical capacity is finite (for the same reason
of the finiteness of the capacity of the perceptron where one exactly
has $\Delta_{2}-\Delta=1-q$). Contrary to the sign activation (where
the effective parameters are such that $\Delta_{2}-\Delta\simeq\delta\Delta\sqrt{1-q}$),
the ReLU activation function happens to belong to this class (with
$\delta\Delta=\frac{1}{2}$). The RS estimate of the critical capacity
is therefore given by 
\begin{equation}
\alpha_{c}^{\text{RS}}=\frac{2\delta\Delta}{\Delta_{2}-\Delta_{-1}}\,.
\end{equation}
One correctly recovers $\alpha_{c}=2$ in the case of the perceptron~\citep{gardner1988The}
whereas for the committee machine with ReLU activation one has $\alpha_{c}=2\left(1-\frac{1}{\pi}\right)^{-1}\simeq2.934$.
As for the sign activation, one expects also for the ReLU activation
that RS is unstable before $\alpha_{c}$. Indeed we have computed
the stability of the RS solution in the large $K$ limit and we found
$\alpha_{\text{AT}}\simeq1.721$ which is far smaller than the corresponding
value of the sign activation. This suggests that strong RSB effects
are at play.

We have therefore used a 1RSB ansatz to better estimate the critical
capacity in the ReLU case. This can be obtained by taking the limits
$q_{1}\to1$ for intra-block overlap and $m\to0$ for the Parisi parameter.
We found $\alpha_{c}^{\text{1RSB}}\simeq2.6643$\footnote{In a previous version of the paper a value of $\alpha_{c}^{\text{1RSB}}\simeq2.92$ was reported. After the work of~\citep{zavatone2020activation} we found that it was incorrect, due to a bad initialization of the solver of the saddle point equations, that let the order parameter converge on the RS saddle point.}, which is not too far
from the RS result.

For the sake of brevity, we mention that the results on the non divergent
capacity with $K$ generalize to other monotone smooth functions such
as the sigmoid.

\noindent \textit{Typical distances}. In order to get a quantitative
understanding of the geometrical structure of the weight space, we
have also derived the so called Franz-Parisi entropy for the committee
machine with a generic transfer function. This framework was originally
introduced in~\citep{franz1995recipes} to study the metastable states
of mean-field spin glasses and only recently it was used to study
the landscape of the solutions of the perceptron~\citep{huang2014origin}.
The basic idea is to sample a solution $\tilde{W}$ from the equilibrium
Boltzmann measure and to study the entropy landscape around it. In
the binary setting, it turns out that the equilibrium solutions of
the learning problem are isolated; this means that, for any positive
value of $\alpha$, one must flip an extensive number of weights to
go from an equilibrium solution to another one. The Franz-Parisi entropy
is defined as 
\begin{equation}
\mathcal{F}_{\text{FP}}\left(S\right)=\frac{1}{N}\left\langle \!\frac{\int\!\mathrm{d}\mu\left(\tilde{W}\right)\,\mathbb{X}_{\xi,\sigma}\left(\tilde{W}\right)\ln\mathcal{N}_{\xi,\sigma}\!\left(\tilde{W},S\right)}{\int\!\mathrm{d}\mu\left(\tilde{W}\right)\,\mathbb{X}_{\xi,\sigma}\left(\tilde{W}\right)}\!\right\rangle _{\!\xi,\sigma}\label{FP}
\end{equation}
where $\mathcal{N}_{\xi,\sigma}\!\left(\tilde{W},S\right)\!=\!\int\mathrm{d}\mu\!\left(W\right)\mathbb{X}_{\xi,\sigma}\!\left(W\right)\prod_{l=1}^{K}\delta\!\left(W_{l}\cdot\tilde{W}_{l}-\frac{N}{K}S\right)$
counts the number of solutions at a distance $d=\frac{1-S}{2}$ from
a reference $\tilde{W}$. The distance constraint is imposed by fixing
the overlap between every sub-perceptron of $W$ and $\tilde{W}$
to $\frac{S}{K}$. The quantity defined in eq.~(\ref{FP}) can again
be computed by the replica method. However, the expression for $K$
finite is quite difficult to analyze numerically since the energetic
term contains $4K$ integrals. The large $K$ limit is instead easier
and, again, the only difference with the perceptron expression is
that the order parameters are replaced with effective ones in the
energetic term (see appendix~\ref{sec:Franz-Parisi-potential} for
details).

\begin{figure}
\begin{centering}
\includegraphics[width=0.48\columnwidth]{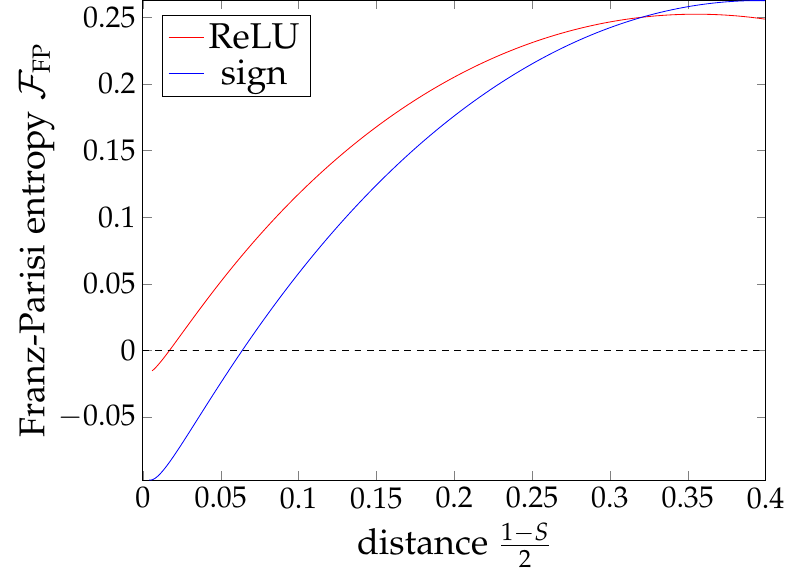} \includegraphics[width=0.48\columnwidth]{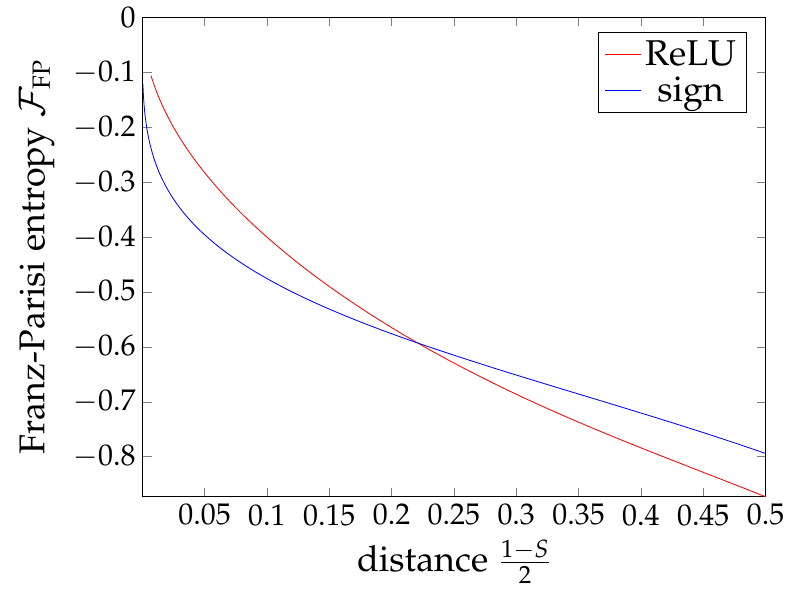}
\par\end{centering}
\caption{(Left panel) Typical solutions are isolated in 2 layer neural networks
with binary weights. We plot the Franz-Parisi entropy $\mathcal{F}_{\text{FP}}$
as a function of the distance from a typical reference solution $\tilde{W}$
for the committee machine with ReLU activations (red line) and sign
activations (blue line) in the limit of large number of neurons in
the hidden layer $K$. We have used $\alpha=0.6$. For both curves
there is a value of the distance for which the entropy becomes negative.
This signals that typical solutions are isolated. (Right panel) Franz-Parisi
entropy as a function of distance, normalized with respect to the
unconstrained $\alpha=0$ case, for spherical weights and for $\alpha=1$.}
\label{Fig::FP}
\end{figure}

In Fig.~\ref{Fig::FP} we plot the Franz-Parisi entropy $\mathcal{F}_{\text{FP}}$
as a function of the distance $d=\frac{1-S}{2}$ from a typical reference
solution $\tilde{W}$ for the committee machine with both sign and
ReLU activations. The numerical analysis shows that, as in the case
of the binary perceptron, also in the 2-layer case solutions are isolated
since there is a minimal distance $d^{*}$ below which the entropy
becomes negative. This minimal distance increases with the constraint
density $\alpha$. However at a given value of $\alpha$, typical
solutions of the committee machine with ReLU activations are less
isolated than the ones of the sign counterpart. The same framework
applies to the case of spherical weights where we find that the minimum
distance between typical solutions is smaller for the ReLU case.

\begin{figure}
\begin{centering}
\includegraphics[width=0.48\columnwidth]{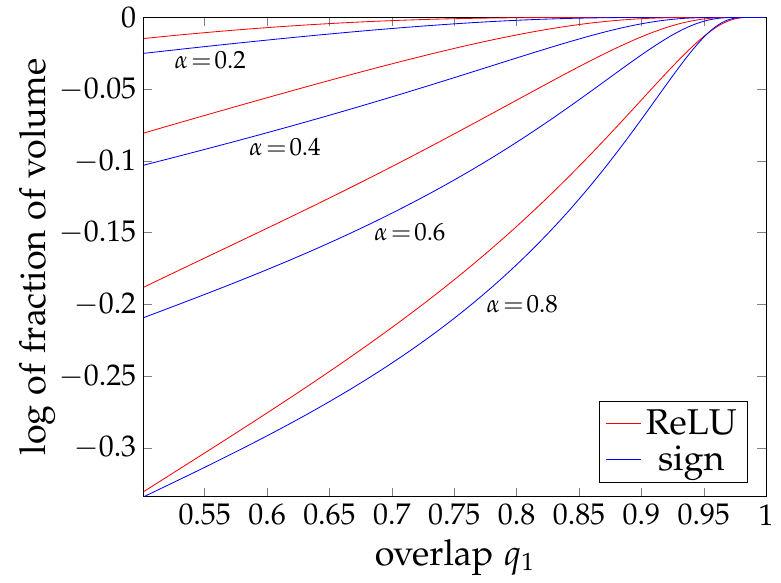} \includegraphics[width=0.48\columnwidth]{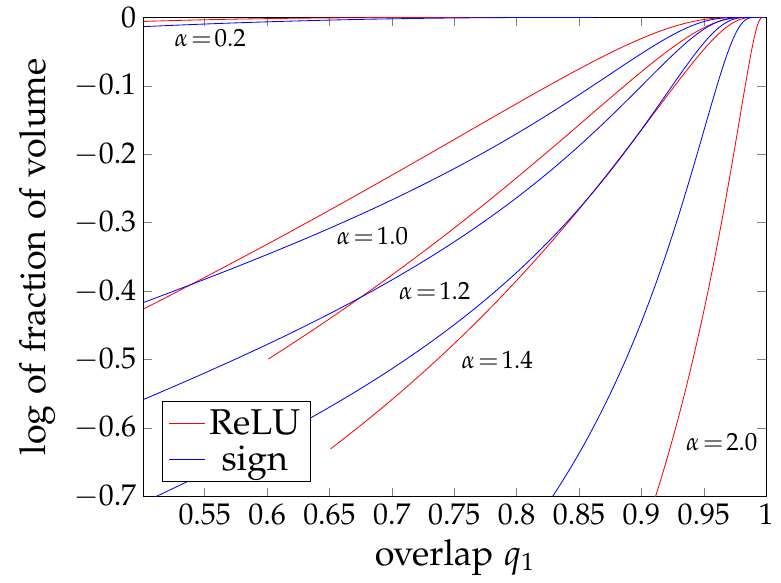}
\par\end{centering}
\caption{Numerical evidence of the greater robustness of the minima of ReLU
transfer function (red line) compared with the sign one (blue line)
using the large deviation analysis in the binary (left panel) setting
and spherical setting (right panel). The exchange in the curves in
both settings for $\alpha$ sufficiently large is due to the fact
that the algorithmic threshold of the ReLU is reached before the corresponding
one of the sign case.}
\label{Fig::LD}
\end{figure}

\noindent \textit{Large deviation analysis}. The results of the previous
section show that the Franz-Parisi framework does not capture the
features of high local entropy regions. These regions indeed exist
since algorithms can be observed to find solutions belonging to large
connected clusters of solutions. In order to study the properties
of wide flat minima or high local entropy regions one needs to introduce
a large deviation measure~\citep{baldassi2015subdominant}, which
favors configurations surrounded by an exponential number of solutions
at small distance. This amounts to study a system with $y$ real replicas
constrained to be at a distance $d$ from a reference configuration
$\tilde{W}$. The high local entropy region is found around $d\simeq0$
in the limit of large $y$. As shown in~\citep{baldassi2019shaping},
an alternative approach can be obtained by directly constraining the
set of $y$ replicas to be at a given mutual distance $d$, that is:
\begin{eqnarray}
Z_{\text{LD}}\left(d,y\right) & = & \int\!\prod_{a=1}^{y}\mathrm{d}\mu\left(W^{a}\right)\prod_{a=1}^{y}\mathbb{X}_{\xi,\sigma}\left(W^{a}\right)\nonumber \\
 &  & \times\prod_{\substack{a<b\\
l
}
}\delta\left(W_{l}^{a}\cdot W_{l}^{b}-\frac{N\left(1-2d\right)}{K}\right)
\end{eqnarray}
This last approach has the advantage of simplifying the calculations,
since it is related to the standard 1RSB approach on the typical Gardner
volume~\citep{krauth-mezard} given in equation~(\ref{GardnerVolume}):
the only difference is that the Parisi parameter $m$ and the intra-block
overlap $q_{1}$ are fixed as external parameters, and play the same
role of $y$ and $1-2d$ respectively. Therefore $m$ is not limited
anymore to the standard range $\left[0,1\right]$; indeed, the large
$m$ regime is the significant one for capturing high local entropy
regions. In the large $m$ and $K$ limit the large deviation free
entropy $\mathcal{F}_{\text{LD}}\equiv\frac{1}{N}\left\langle \ln Z_{\text{LD}}\right\rangle _{\xi,\sigma}$
reads 
\begin{equation}
\begin{split}\mathcal{F}_{\text{LD}}\left(q_{1}\right)=\mathcal{G}_{S}\left(q_{1}\right)+\alpha\,\mathcal{G}_{E}\left(q_{1}\right)\end{split}
\end{equation}
where, again, the entropic term has a different expression depending
on the constraint over the weights $W$. Its expression, together
with the corresponding energetic term, is reported in appendix~\ref{sec:Large-deviation-analysis}.

We report in Fig.~\ref{Fig::LD} the numerical results for both binary
and spherical weights of the large deviation entropy (normalized with
respect to the unconstrained $\alpha=0$ case) as a function of $q_{1}$.
For both sign and ReLU activations, the region $q_{1}\simeq1$ is
flat around zero. This means that there exist $\tilde{W}$ references
around which the landscape of solutions is basically indistinguishable
from the $\alpha=0$ case where all configurations are solutions.
We also find that the ReLU curve, in the vicinity of $q_{1}\simeq1$,
is always more entropic than the corresponding one of the sign. This
picture is valid for sufficiently low values of $\alpha$; for $\alpha$
greater of a certain value $\alpha^{*}$ the two curves switch. This
is due to the fact that the two models have completely different critical
capacities (divergent in the sign case, finite in the ReLU case) so
that one expects that clusters of solutions disappear at a smaller
constrained capacity when ReLU activations are used.

\begin{figure}
\begin{centering}
\includegraphics[width=0.48\columnwidth]{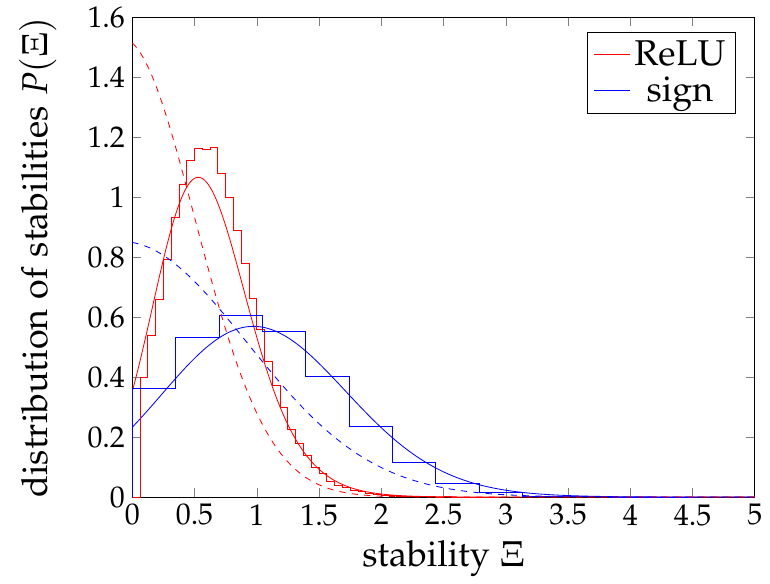}
\includegraphics[width=0.48\columnwidth]{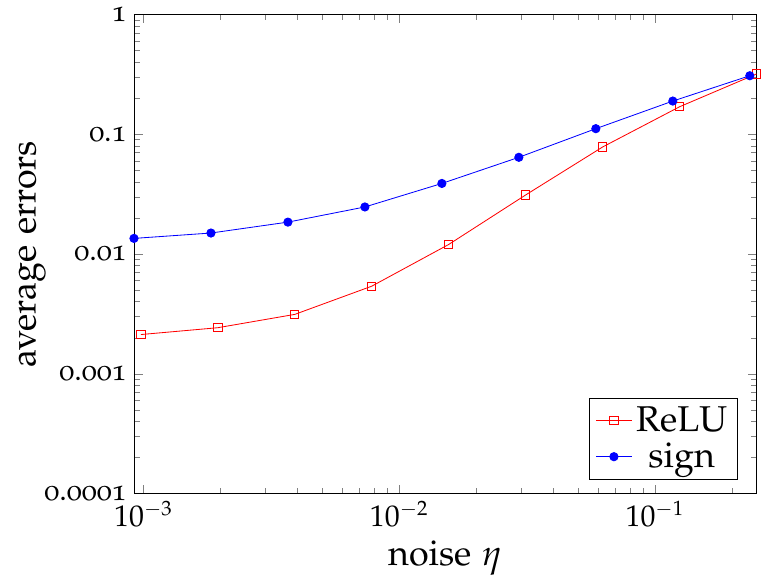}
\par\end{centering}
\caption{(Left panel) Dashed lines: theoretical stability curves for the typical
solutions, for binary weights at $\alpha=0.4$. Solid lines: comparison
between the numerical and theoretical stability distributions in the
large deviation scenario, same $\alpha$. We have used $q_{1}=0.85$,
$y=20$. (Right panel) Robustness of the reference configuration found
by replicated simulated annealing when one pattern is perturbed by
flipping a certain fraction $\eta$ of entries.}

\label{Fig::stability}
\end{figure}

\noindent \textit{Stability distribution and robustness}. To corroborate
our previous results, we have also computed (see appendix~\ref{sec:Distribution-of-stabilities}),
for various models and types of solutions $W$, the distribution of
the \emph{stabilities} $\Xi=\left\langle \frac{\sigma}{\sqrt{K}}\sum_{l=1}^{K}c_{l}\,\tau_{l}^{\mu}\right\rangle _{\xi,\sigma}$,
which measure the distance from the threshold at the output unit in
the direction of the correct label $\sigma$, cf.~eq.~(\ref{eq:out}).
Previous calculations~\citep{kepler1988domains} have shown that
in the simple case of the spherical perceptron at the critical capacity
the stability distribution around a typical solution $W$ develops
a delta peak in the origin $\Xi\simeq0$; we confirmed that even in
the two-layer case the stability distribution of a typical solution,
being isolated, also has its mode at $\Xi\simeq0$ even at lower $\alpha$,
see the dashed lines in Fig.~\ref{Fig::stability} (left). A solution
surrounded by an exponential number of other solutions, instead, should
be more robust and be centered away from $0$. Our calculations show
that this is indeed the case both for the sign and for the ReLU activations,
and we have confirmed the results by numerical simulations. In Fig.~\ref{Fig::stability}
(left) we show the analytical and numerical results for the case of
binary weights at $\alpha=0.4$ with $y=20$ replicas at $q_{1}=0.85$.
For the numerical results, we have used simulated annealing on a system
with $K=32$ ($K=33$) for the ReLU (sign) activations (respectively),
and $N=K^{2}\simeq10^{3}$. We have simulated a system of $y$ interacting
replicas that is able to sample from the local-entropic measure~\citep{baldassi_unreasonable_2016}
with the RRR Monte Carlo method~\citep{baldassi2017method}, ensuring
that the annealing process was sufficiently slow such that at the
end of the simulation all replicas were solutions, and controlling
the interaction such that the average overlap between replicas was
equal to $q_{1}$ within a tolerance of $0.01$. The results were
averaged over $20$ samples. As seen in Fig.~\ref{Fig::stability}
(left), the agreement with the analytical computations is remarkable,
despite the small values of $\nicefrac{N}{K}$ and $K$ and the approximations
introduced by sampling with simulated annealing.

The stabilities for the sign and ReLU activations are qualitatively
similar, but quantitatively we observe that in all cases the curves
for the ReLU case have a peak closer to $0$ and a smaller variance.
These are not, however, directly comparable, and it is difficult to
tell from the stability curves alone which choice is more robust.
We have thus directly measured, on the results of the simulations,
the effect of introducing noise in the input patterns. For each trained
group of $y$ replicas, we used the configuration of the reference
$\tilde{W}$ (which lays in the middle of the cluster of solutions)
and we measured the probability that a pattern of the training set
would be misclassified if perturbed by flipping a fraction $\eta$
of randomly chosen entries. We explored a wide range of values of
the noise $\eta$ and sampled $50$ perturbations per pattern. The
results are shown in Fig.~\ref{Fig::stability} (right), and they
confirm that the networks with ReLU activations are more robust than
those with sign activations for this $\alpha$, in agreement with
the results of Fig.~\ref{Fig::LD}. We also verified that the reference
configuration is indeed more robust than the individual replicas.
The results for other choices of the parameters are qualitatively
identical. Our preliminary tests show that the same phenomenology
is maintained when the network architecture is changed to a fully-connected
scheme, in which each hidden unit is connected to all of the input
units.

The architecture of the model that we have analyzed here is certainly
very simplified compared to state of the art deep neural networks
used in applications. Investigating deeper models would certainly
be of great interest, but extremely challenging with current analytical
techniques, and is thus an open problem. Extending our analysis to
a one-hidden-layer fully-connected model, on the other hand, would
in principle be feasible (the additional complication comes from the
permutation symmetry of the hidden layer). However, based on the existing
literature (e.g.~\citep{barkai1992broken,urbanczik1997storage}),
and our preliminary numerical experiments mentioned above, we do not
expect that such extension would result in qualitatively different
outcomes compared to our tree-like model.

\noindent \bibliographystyle{unsrturl}
\bibliography{references-ReLU}

\onecolumngrid

\appendix

\section{Model}

Our model is a tree-like committee machine with $N$ weights $W_{li}$
divided into $K$ groups of $\nicefrac{N}{K}$ entries. We use the
index $l=1,\dots,K$ for the group and $i=1,\dots,\frac{N}{K}$ for
the entry. We consider two cases, the binary case $W_{li}=\pm1$for
all $l,i$ and the continuous case with spherical constraints on each
group, $\sum_{i=1}^{N/K}W_{li}^{2}=N/K$ for all $l$.

The training set consists of $p=\alpha N$ random binary i.i.d. patterns.
The inputs are denoted by $\xi_{li}^{\mu}$ and the outputs by $\sigma^{\mu}$,
where $\mu=1,\dots,p$ is the pattern index.

In our analysis, we write the model using a generic activation function
$g\left(x\right)$ for the first layer (hidden) units. We will consider
two cases: the sign case $g\left(x\right)=\mathrm{sgn}$$\left(x\right)$
and the ReLU case $g\left(x\right)=\max\left(0,x\right)$. The output
of any given unit $l$ in response to a pattern $\mu$ is thus written
as
\begin{equation}
\tau_{l}^{\mu}=g\left(\frac{1}{\sqrt{N/K}}\sum_{i}W_{li}\xi_{li}^{\mu}\right)
\end{equation}
The connection weights between the first layer and the output are
denoted by $c_{l}$ and considered binary and fixed; for the case
of ReLU activations we set the first half $l=1,\dots,K/2$ to the
value $+1$ and the rest to $-1$; for the case of the sign activations
we can set them to all to $+1$ without loss of generality.

A configuration of the weights $W$ solves the training problem if
it classifies correctly all the patterns; we denote this with the
indicator function

\begin{equation}
\mathbb{X}_{\xi,\sigma}\left(W\right)=\prod_{\mu}\Theta\left(\frac{\sigma^{\mu}}{\sqrt{K}}\sum_{l}c_{l}\,\tau_{l}^{\mu}\right)
\end{equation}
where $\Theta\left(x\right)=1$ if $x>0$ and $0$ otherwise is the
Heaviside step function.

The volume of the space of configurations that correctly classify
the whole training set is then
\begin{equation}
Z=\int\mathrm{d}\mu\left(W\right)\mathbb{X}_{\xi,\sigma}\left(W\right)
\end{equation}
where $\mathrm{d}\mu\left(W\right)$ is the flat measure over the
admissible values of the $W$, depending on whether we're analyzing
the binary or the spherical case. The average of the log-volume over
the distribution of the patterns, $\left\langle \log Z\right\rangle _{\xi,\sigma}$,
is the free entropy of the model $\mathcal{F}$. We can evaluate it
in the large $N$ limit with the the ``replica trick'', using the
formula $\log Z=\lim_{n\to0}\partial_{n}Z^{n}$, computing the average
for integer $n$ and then taking the limit $n\to0$.

As explained in the main text, in all cases the resulting expression
takes the form
\begin{equation}
\mathcal{F}=\mathcal{G}_{S}+\alpha\mathcal{G}_{E}
\end{equation}
where the $\mathcal{G}_{S}$ part is only affected by the spherical
or binary nature of the weights, whereas the $\mathcal{G}_{E}$ part
is only affected by $K$ and by the activation function $g$. Determining
their value requires to compute a saddle-point over some overlap parameters
$q_{l}^{ab}$ with $a,b=1,\dots,n$ representing overlaps between
replicas, and their conjugates $\hat{q}_{l}^{ab}$; in turn, this
requires an ansatz about the structure of the saddle-point in order
to perform the $n\to0$ limit.

For the spherical weights, the $\mathcal{G}_{S}$ part (before the
$n\to0$ limit) reads
\begin{equation}
\mathcal{G}_{S}^{\mathrm{sph}}=-\frac{1}{2nK}\sum_{l}\sum_{a,b}q_{l}^{ab}\hat{q}_{l}^{ab}+\frac{1}{nK}\sum_{l}\log\int\prod_{a}\mathrm{d}W^{a}\ e^{\frac{1}{2}\sum_{a,b}\hat{q}_{l}^{ab}W^{a}W^{b}}
\end{equation}
while for the binary case we have a very similar expression, except
that the summations don't have the $a=b$ case and the integral over
$W^{a}$ becomes a summation:
\begin{equation}
\mathcal{G}_{S}^{\mathrm{bin}}=-\frac{1}{2nK}\sum_{l=1}^{K}\sum_{a\ne b}q_{l}^{ab}\hat{q}_{l}^{ab}+\frac{1}{nK}\sum_{l=1}^{K}\log\sum_{W^{a}=\pm1}\ e^{\frac{1}{2}\sum_{a\ne b}\hat{q}_{l}^{ab}W^{a}W^{b}}
\end{equation}
The $\mathcal{G}_{E}$ part reads:
\begin{align}
\mathcal{G}_{E} & =\frac{1}{n}\mathbb{E}_{\sigma}\log\int\prod_{la}\frac{\mathrm{d}u_{l}^{a}\mathrm{d}\hat{u}_{l}^{a}}{2\pi}\prod_{a}\ \Theta\left(\sigma^{\mu}\frac{1}{\sqrt{K}}\sum_{l}c_{l}g\left(u_{l}^{a}\right)\right)e^{-\sum_{al}i\hat{u}_{l}^{a}u_{l}^{a}-\frac{1}{2}\sum_{l}\sum_{ab}q_{l}^{ab}\hat{u}_{l}^{a}\hat{u}_{l}^{b}}
\end{align}
In all cases, we study the problem in the large $K$ limit, which
allows to invoke the central limit theorem and leads to a crucial
simplification of the expressions.

\section{Critical capacity}

\subsection{Replica symmetric ansatz\label{subsec:RS}}

In the replica-symmetric (RS) case we seek solutions of the form $q_{l}^{ab}=\delta_{ab}+\left(1-\delta_{ab}\right)q$
for all $l,a,b$, where $\delta_{ab}$ is the Kronecker delta symbol,
and similarly for the conjugated parameters, $\hat{q}_{l}^{ab}=\delta_{ab}\hat{Q}+\left(1-\delta_{ab}\right)\hat{q}$
for all $l,a,b$. The resulting expressions, as reported in the main
text, are:
\begin{align}
\mathcal{G}_{S}^{\mathrm{sph}} & =\frac{1}{2}\hat{Q}+\frac{1}{2}q\hat{q}+\frac{1}{2}\ln\frac{2\pi}{\hat{Q}+\hat{q}}+\frac{\hat{q}}{2\left(\hat{Q}+\hat{q}\right)}\\
\mathcal{G}_{S}^{\mathrm{bin}} & =-\frac{\hat{q}}{2}\left(1-q\right)+\int Dz\,\ln2\cosh\left(z\sqrt{\hat{q}}\right)\\
\mathcal{G}_{E} & =\int Dz_{0}\,\ln H\left(-\sqrt{\frac{\Delta-\Delta_{-1}}{\Delta_{2}-\Delta}}z_{0}\right)
\end{align}
where $Dz\equiv\mathrm{dz}\frac{e^{-\frac{1}{2}z^{2}}}{\sqrt{2\pi}}$
is a Gaussian measure, $H\left(x\right)=\int_{x}^{\infty}Dz=\frac{1}{2}\mathrm{erfc}\left(\frac{x}{\sqrt{2}}\right)$
and the expressions of $\Delta$, $\Delta_{2}$ and $\Delta_{-1}$
depend on the activation function $g$:
\[
\Delta_{-1}^{\mathrm{sgn}}=0;\qquad\qquad\qquad\Delta^{\mathrm{sgn}}=1-\frac{2}{\pi}\mathrm{arccos}\left(q\right);\qquad\qquad\qquad\Delta_{2}^{\mathrm{sgn}}=1
\]

\[
\Delta_{-1}^{\mathrm{ReLU}}=\frac{1}{2\pi};\qquad\Delta^{\mathrm{ReLU}}=\frac{\sqrt{1-q^{2}}}{2\pi}+\frac{q}{\pi}\mathrm{arctan}\sqrt{\frac{1+q}{1-q}};\qquad\Delta_{2}^{\mathrm{ReLU}}=\frac{1}{2}
\]
The values of the overlaps and conjugated parameters are found by
setting to $0$ the derivatives of the free entropy.

The critical capacity $\alpha_{c}$ is found in the binary case by
seeking numerically the value of $\alpha$ for which the saddle point
solutions returns a zero free entropy.For the spherical case, instead,
$\alpha_{c}$ is determined by finding the value of $\alpha$ such
that $q\to1$, which can be obtained analytically by reparametrizing
$q=1-\delta q$ and expanding around $\delta q\ll1$. In this limit,
we must also reparametrize $\Delta$ using $\Delta_{2}-\Delta=\delta\Delta\,\delta q^{x}$,
where $x$ is an exponent that depends on the activation function:
it is $x=\nicefrac{1}{2}$ for the sign and $x=1$ for the ReLU. Due
to this difference in this exponent, $\alpha_{c}$ diverges in the
sign activation case (as was shown in \citep{barkai1992broken,engel1992storage}),
while for the ReLU activations it converges to $2\left(1-\frac{1}{\pi}\right)^{-1}$.
However, the RS result for the spherical case is only an upper bound,
and a more accurate result requires replica-symmetry breaking.

\subsection{1RSB ansatz\label{subsec:1RSB-ansatz}}

In the one-step replica-symmetry-breaking ($1$-RSB) ansatz we seek
solutions with 3 possible values of the overlaps $q_{l}^{ab}$ and
their conjugates. We group the $n$ replicas in $\frac{n}{m}$ groups
of $m$ replicas each, and denote with $q_{0}$ the overlaps among
different groups and with $q_{1}$ the overlaps within the same group.
As before, the self overlap is $q^{aa}=1$ and its conjugate $\hat{q}^{aa}=\hat{Q}$
(these are only relevant in the spherical case).

The resulting expressions are:
\begin{align}
\mathcal{G}_{S}^{\mathrm{sph}} & =\frac{1}{2}\hat{Q}+\frac{1}{2}\hat{q}_{1}q_{1}+\frac{m}{2}\left(q_{0}\hat{q}_{0}-q_{1}\hat{q}_{1}\right)+\frac{1}{2}\left[\ln\frac{2\pi}{\hat{Q}+\hat{q}_{1}}+\frac{1}{m}\ln\left(\frac{\hat{Q}+\hat{q}_{1}}{\hat{Q}+\hat{q}_{1}-m\left(\hat{q}_{1}-\hat{q}_{0}\right)}\right)+\frac{\hat{q}_{0}}{\hat{Q}+\hat{q}_{1}-m\left(\hat{q}_{1}-\hat{q}_{0}\right)}\right]\label{eq:Gs_sph}\\
\mathcal{G}_{S}^{\mathrm{bin}} & =-\frac{\hat{q}_{1}}{2}\left(1-q_{1}\right)+\frac{m}{2}\left(q_{0}\hat{q}_{0}-q_{1}\hat{q}_{1}\right)+\frac{1}{m}\int Du\,\ln\int Dv\left(2\cosh\left(\sqrt{\hat{q}_{0}}u+\sqrt{\hat{q}_{1}-\hat{q}_{0}}v\right)\right)^{m}\label{eq:Gs_bin}\\
\mathcal{G}_{E} & =\frac{1}{m}\int Dz_{0}\,\ln\int Dz_{1}H\left(-\frac{\sqrt{\Delta_{0}-\Delta_{-1}}z_{0}+\sqrt{\Delta_{1}-\Delta_{0}}z_{1}}{\sqrt{\Delta_{2}-\Delta_{1}}}\right)^{m}\label{eq:Ge}
\end{align}

the expressions of $\Delta_{2}$ and $\Delta_{-1}$ are the same as
in the RS case. The expressions of $\Delta_{0}$ and $\Delta_{1}$
take the same form as the RS expressions for $\Delta$, except that
$q_{0}$ and $q_{1}$ must be used instead of $q$.

Similarly to the RS case, in order to determine the critical capacity
$\alpha_{c}$ in the spherical case, we need to find the value of
$\alpha$ such that $q_{1}\to1$. In this case however we must also
have $m\to0$. The scaling is such that $\tilde{m}=m/\left(1-q_{1}\right)$
is finite. The final expression reads

\begin{equation}
\alpha_{c}=\frac{\ln\left(1+\tilde{m}\left(1-q_{0}\right)\right)+\frac{\tilde{m}q_{0}}{1+\tilde{m}\left(1-q_{0}\right)}}{2f\left(q_{0},\tilde{m}\right)}
\end{equation}
where
\begin{equation}
f\left(q_{0},\tilde{m}\right)=\int Dz_{0}\,\ln\left[\frac{\sqrt{\delta\Delta}\,e^{-\frac{\left(\Delta_{0}-\Delta_{-1}\right)\tilde{m}}{\delta\Delta+\left(\Delta_{2}-\Delta_{0}\right)\tilde{m}}\frac{z_{0}^{2}}{2}}}{\sqrt{\delta\Delta+\left(\Delta_{2}-\Delta_{0}\right)\tilde{m}}}H\left(-\sqrt{\frac{\Delta_{0}-\Delta_{-1}}{\Delta_{2}-\Delta_{0}}}\frac{\sqrt{\delta\Delta}\,z_{0}}{\sqrt{\delta\Delta+\left(\Delta_{2}-\Delta_{0}\right)\tilde{m}}}\right)+H\left(\sqrt{\frac{\Delta_{0}-\Delta_{-1}}{\Delta_{2}-\Delta_{0}}}z_{0}\right)\right]
\end{equation}
The expression of $\delta\Delta$ is the same as in the RS case using
$\Delta_{1}$ instead of $\Delta$; the values of $q_{0}$ and $\tilde{m}$
are determined by saddle point equations as usual.

\section{Franz-Parisi potential\label{sec:Franz-Parisi-potential}}

\noindent The Franz-Parisi entropy \citep{franz1995recipes,huang2014origin}
is defined as (cf.~eq.~(\ref{FP}) of the main text):
\begin{equation}
\mathcal{F}_{\text{FP}}\left(S\right)\!=\!\frac{1}{N}\left\langle \!\frac{\int\!\mathrm{d}\mu\left(\tilde{W}\right)\,\mathbb{X}_{\xi,\sigma}\left(\tilde{W}\right)\ln\mathcal{N}_{\xi,\sigma}\!\left(\tilde{W},S\right)}{\int\!\mathrm{d}\mu\left(\tilde{W}\right)\,\mathbb{X}_{\xi,\sigma}\left(\tilde{W}\right)}\!\right\rangle _{\!\xi,\sigma}
\end{equation}
where $\mathcal{N}_{\xi,\sigma}\!\left(\tilde{W},S\right)\!=\!\int\mathrm{d}\mu\!\left(W\right)\mathbb{X}_{\xi,\sigma}\!\left(W\right)\prod_{l=1}^{K}\delta\!\left(W_{l}\cdot\tilde{W}_{l}-\frac{N}{K}S\right)$
counts the number of solutions at a distance $\frac{1-S}{2}$ from
the reference $\tilde{W}$. In this expression, $\tilde{W}$ represents
a typical solution. The evaluation of this quantity requires the use
of two sets of replicas: $n$ replicas of the reference configuration
$\tilde{W}$, for which we use the indices $a$ and $b$, and $r$
replicas of the surrounding configurations $W$, for which we use
the indices $c$ and $d$. The computation proceeds following standard
steps, leading to these expressions, written in terms of the overlaps
$q_{l}^{ab}=\frac{K}{N}\sum_{i=1}^{N/K}\tilde{W}_{li}^{a}\tilde{W}_{li}^{b}$,
$p_{l}^{cd}=\frac{K}{N}\sum_{i=1}^{N/K}W_{li}^{c}W_{li}^{d}$, $t_{l}^{ac}=\frac{K}{N}\sum_{i=1}^{N/K}\tilde{W}_{li}^{a}W_{li}^{c}$
and their conjugate quantities:

\noindent 
\begin{equation}
\begin{split}\mathcal{F}_{\text{FP}}=\mathcal{G}_{S}+\alpha\,\mathcal{G}_{E}\end{split}
\end{equation}
\begin{eqnarray}
\mathcal{G}_{S}^{\mathrm{sph}} & = & -\frac{1}{2nK}\sum_{l=1}^{K}\sum_{a,b}q_{l}^{ab}\hat{q}_{l}^{ab}-\frac{1}{2nK}\sum_{l=1}^{K}\sum_{c,d}p_{l}^{cd}\hat{p}_{l}^{cd}-\frac{1}{nK}\sum_{l=1}^{K}\sum_{a>1,c}t_{l}^{ac}\hat{t}_{l}^{ac}-\frac{S}{nK}\sum_{l=1}^{K}\sum_{a>1,c}\hat{S}_{l}^{c}+\\
 &  & +\frac{1}{nK}\ln\int\prod_{al}\mathrm{d}\tilde{W}_{l}^{a}\prod_{cl}\mathrm{d}W_{l}^{c}\prod_{l}e^{\frac{1}{2}\sum_{a,b}\hat{q}_{l}^{ab}\tilde{W}_{l}^{a}\tilde{W}_{l}^{b}+\frac{1}{2}\sum_{c,d}\hat{p}_{l}^{cd}W_{l}^{c}W_{l}^{d}+\sum_{a>1,c}\hat{t}_{l}^{ac}\tilde{W}_{l}^{a}W_{l}^{c}+\tilde{W}_{l}^{a=1}\sum_{c}\hat{S}_{l}^{c}W_{l}^{c}}\nonumber \\
\mathcal{G}_{S}^{\mathrm{bin}} & = & -\frac{1}{2nK}\sum_{l=1}^{K}\sum_{a\ne b}q_{l}^{ab}\hat{q}_{l}^{ab}-\frac{1}{2nK}\sum_{l=1}^{K}\sum_{c\ne d}p_{l}^{cd}\hat{p}_{l}^{cd}-\frac{1}{nK}\sum_{l=1}^{K}\sum_{a>1,c}t_{l}^{ac}\hat{t}_{l}^{ac}-\frac{S}{nK}\sum_{l=1}^{K}\sum_{a>1,c}\hat{S}_{l}^{c}+\\
 &  & +\frac{1}{nK}\ln\sum_{\tilde{W}_{l}^{a}=\pm1}\sum_{W_{l}^{c}=\pm1}\prod_{l}e^{\frac{1}{2}\sum_{a\ne b}\hat{q}_{l}^{ab}\tilde{W}_{l}^{a}\tilde{W}_{l}^{b}+\frac{1}{2}\sum_{c\ne d}\hat{p}_{l}^{cd}W_{l}^{c}W_{l}^{d}+\sum_{a>1,c}\hat{t}_{l}^{ac}\tilde{W}_{l}^{a}W_{l}^{c}+\tilde{W}_{l}^{a=1}\sum_{c}\hat{S}_{l}^{c}W_{l}^{c}}\nonumber \\
\mathcal{G}_{E} & = & \frac{1}{n}\mathbb{E}_{\sigma}\ln\int\prod_{la}\frac{\mathrm{d}\lambda_{l}^{a}\mathrm{d}\hat{\lambda}_{l}^{a}}{2\pi}\prod_{lc}\frac{\mathrm{d}u_{l}^{c}\mathrm{d}\hat{u}_{l}^{c}}{2\pi}e^{i\lambda_{l}^{a}\hat{\lambda}_{l}^{a}+iu_{l}^{c}\hat{u}_{l}^{c}}\prod_{a}\theta\left(\frac{\sigma}{\sqrt{K}}\sum_{l}c_{l}\,g\left(\lambda_{l}^{a}\right)\right)\prod_{c}\theta\left(\frac{\sigma}{\sqrt{K}}\sum_{l}c_{l}\,g\left(u_{l}^{c}\right)\right)\times\\
 &  & \times\prod_{l}e^{-\frac{1}{2}\sum_{a}\left(\hat{\lambda}_{l}^{a}\right)^{2}-\frac{1}{2}\sum_{c}\left(\hat{u}_{l}^{c}\right)^{2}-\sum_{a<b}q_{l}^{ab}\hat{\lambda}_{l}^{a}\hat{\lambda}_{l}^{b}-\sum_{c<d}p_{l}^{cd}\hat{u}_{l}^{c}\hat{u}_{l}^{d}-\sum_{a>1,c}t_{l}^{ac}\hat{\lambda}_{l}^{a}\hat{u}_{l}^{c}-S\hat{\lambda}_{l}^{a=1}\sum_{c}\hat{u}_{l}^{c}}\nonumber 
\end{eqnarray}
The calculation proceeds by taking the RS ansatz with the same structure
as that of sec.~\ref{subsec:RS} and the limit $n\to0$, $r\to0$.
We obtain, in the large $K$ limit:

\begin{eqnarray}
\mathcal{G}_{S}^{\mathrm{\mathrm{sph}}} & = & \frac{1}{2}\hat{P}+\frac{1}{2}\hat{p}p+\hat{t}t-\hat{S}S+\frac{1}{2}\ln\frac{2\pi}{\hat{P}+\hat{p}}+\frac{1}{\hat{P}+\hat{p}}\left[\frac{\hat{p}}{2}+\frac{\left(\hat{S}-\hat{t}\right)^{2}\left(\frac{\hat{Q}}{2}+\hat{q}\right)}{\left(\hat{Q}+\hat{q}\right)^{2}}+\frac{\hat{t}\left(\hat{S}-\hat{t}\right)}{\hat{Q}+\hat{q}}\right]\\
\mathcal{G}_{S}^{\mathrm{bin}} & = & -\frac{1}{2}\hat{p}\left(1-p\right)+\hat{t}t-\hat{S}S+\int Du\frac{\sum_{\tilde{W}=\pm1}e^{\tilde{W}\sqrt{\hat{q}}x}\int Dv\ln\left[2\cosh\left(\sqrt{\hat{p}-\frac{\hat{t}^{2}}{\hat{q}}}v+\frac{\hat{t}}{\sqrt{\hat{q}}}u+\left(\hat{S}-\hat{t}\right)\tilde{W}\right)\right]}{2\cosh\left(\sqrt{\hat{q}}u\right)}\\
\mathcal{G}_{E} & = & \int Dz_{0}\frac{\int Dz_{1}H\text{\ensuremath{\left(-\frac{D_{1}z_{1}+\sqrt{\Sigma_{0}\Gamma}z_{0}}{\sqrt{\Sigma_{1}\Gamma-D_{1}^{2}}}\right)} \ensuremath{\ln H\left(-\frac{\sqrt{\Gamma}z_{1}+D_{0}z_{0}/\sqrt{\Sigma_{0}}}{\sqrt{\Delta_{3}}}\right)}}}{H\left(-\sqrt{\frac{\Sigma_{0}}{\Sigma_{1}}}z_{0}\right)}
\end{eqnarray}
where we introduced the auxiliary quantities $\Sigma_{0}$, $\Sigma_{1}$,
$\Gamma$, $D_{0}$, $D_{1}$ and $\Delta_{3}$ which depend on the
choice of the activation function (like the $\Delta_{-1/0/1/2}$ of
the previous section). For the sign activations we get:

\begin{align}
\Delta_{3}^{\mathrm{sgn}} & =\frac{2}{\pi}\arccos p\\
\Sigma_{0}^{\mathrm{sgn}} & =1-\frac{2}{\pi}\arccos q\\
\Sigma_{1}^{\mathrm{sgn}} & =\frac{2}{\pi}\arccos q\\
D_{0}^{\mathrm{sgn}} & =\frac{2}{\pi}\arctan\left(\frac{t}{\sqrt{1-t^{2}}}\right)\\
D_{1}^{\mathrm{sgn}} & =\frac{2}{\pi}\left[\arctan\left(\frac{S}{\sqrt{1-S^{2}}}\right)-\arctan\left(\frac{t}{\sqrt{1-t^{2}}}\right)\right]\\
\Gamma^{\mathrm{sgn}} & =1-\frac{2}{\pi}\arccos p-\frac{\left(D_{0}^{\mathrm{sgn}}\right)^{2}}{\Sigma_{0}^{\mathrm{sgn}}}
\end{align}
For the ReLU activations we get:
\begin{align}
\Delta_{3}^{\mathrm{ReLU}} & =\frac{1}{2}-\frac{\sqrt{1-p^{2}}}{2\pi}-\frac{p}{\pi}\arctan\sqrt{\frac{1+p}{1-p}}\\
\Sigma_{0}^{\mathrm{ReLU}} & =\frac{\sqrt{1-q^{2}}}{2\pi}+\frac{q}{\pi}\arctan\sqrt{\frac{1+q}{1-q}}-\frac{1}{2\pi}\\
\Sigma_{1}^{\mathrm{ReLU}} & =\frac{1}{2}-\frac{\sqrt{1-q^{2}}}{2\pi}-\frac{q}{\pi}\arctan\sqrt{\frac{1+q}{1-q}}\\
D_{0}^{\mathrm{ReLU}} & =\frac{\sqrt{1-t^{2}}}{2\pi}+\frac{t}{\pi}\arctan\sqrt{\frac{1+t}{1-t}}-\frac{1}{2\pi}\\
D_{1}^{\mathrm{ReLU}} & =\frac{\sqrt{1-S^{2}}}{2\pi}+\frac{S}{\pi}\arctan\sqrt{\frac{1+S}{1-S}}-\frac{\sqrt{1-t^{2}}}{2\pi}-\frac{t}{\pi}\arctan\sqrt{\frac{1+t}{1-t}}\\
\Gamma^{\mathrm{ReLU}} & =\frac{\sqrt{1-p^{2}}}{2\pi}+\frac{p}{\pi}\arctan\sqrt{\frac{1+p}{1-p}}-\frac{1}{2\pi}-\frac{\left(D_{0}^{\mathrm{ReLU}}\right)^{2}}{\Sigma_{0}^{\mathrm{ReLU}}}
\end{align}
In order to find the order parameters for any given $\alpha$ and
$S$, we need to set to $0$ the derivatives of the free entropy w.r.t.
the order parameters $q$, $p$, $t$ and the conjugates $\hat{Q}$,
$\hat{q}$, $\hat{P}$, $\hat{p}$, $\hat{t}$, $\hat{S}$, thus obtaining
a system of 9 equations (7 for the binary case) to be solved numerically.
The equations actually reduce to 6 (5 in the binary case) since $q$,
$\hat{Q}$ and $\hat{q}$ are the same ones derived from the typical
case (sec.~\ref{subsec:RS}).

\section{Large deviation analysis\label{sec:Large-deviation-analysis}}

Following~\citep{baldassi2019shaping}, the large deviation analysis
for the description of the high-local-entropy landscape uses the same
equations as the standard $1$RSB expressions eqs.~(\ref{eq:Gs_sph}),
(\ref{eq:Gs_bin}) and~(\ref{eq:Ge}). In this case, however, the
overlap $q_{1}$ is not determined by a saddle point equation, but
rather it is treated as an external parameter that controls the mutual
overlap between the $y$ replicas of the system. Also, the parameter
$m$ is not optimized and it is not restricted to the range $\left[0,1\right]$;
instead, it plays the role of the number of replicas $y$ and it is
generally taken to be large (we normally use either a large integer
number to compare the results with numerical simulations, or we take
the limit $m\to\infty$). For these reason, there are two saddle point
equations less compared to the standard $1$RSB calculation.

The resulting expression for the free entropy $\mathcal{F}_{\text{LD}}\left(q_{1}\right)$
represents, in the spherical case, the log-volume of valid configurations
(solutions at the correct overlap) of the system of $y$ replicas.
These configurations are thus embedded in $\mathcal{S}^{Ky}$ where
$\mathcal{S}$ is the $\nicefrac{N}{K}$-dimensional sphere of radius
$\sqrt{\nicefrac{N}{K}}$. In order to quantify the solution density,
we must normalize $\mathcal{F}_{\text{LD}}\left(q_{1}\right)$, subtracting
the log-volume of all the admissible configurations at a given $q_{1}$
without the solution constraint (which is obtained by the analogous
computation with $\alpha=0$). The resulting quantity is thus upper-bounded
by $0$ (cf.~Fig.~(\ref{Fig::LD}) of the main text). For the binary
case, $\mathcal{F}_{\text{LD}}\left(q_{1}\right)$ is the log of the
number of admissible solutions, and the same normalization procedure
can be applied.

\subsubsection*{Large $m$ limit}

\begin{align}
\mathcal{G}_{S}^{\mathrm{bin}}\left(q_{1}\right) & =-\frac{\hat{q}_{1}}{2}\left(1-q_{1}\right)-\frac{1}{2}\left(\delta q_{0}\,\hat{q}_{1}+q_{1}\delta\hat{q}_{0}\right)+\int Du\,\underset{v}{\max}\left[-\frac{v^{2}}{2}+\ln2\cosh\left(\sqrt{\hat{q}_{1}}u+\sqrt{\delta\hat{q}_{0}}v\right)\right]\\
\mathcal{G}_{S}^{\mathrm{sph}}\left(q_{1}\right) & =\frac{1}{2}\frac{q_{1}}{1-q_{1}+\delta q_{0}}+\frac{1}{2}\text{\ensuremath{\ln}}\left(1-q_{1}\right)\\
\mathcal{G}_{E}\left(q_{1}\right) & =\int Dz_{0}\underset{z_{1}}{\max}\left[-\frac{z_{1}^{2}}{2}+\log H\left(-\frac{\sqrt{\Delta_{1}-\Delta_{-1}}z_{0}+\sqrt{\delta\Delta_{0}}\,z_{1}}{\sqrt{\Delta_{2}-\Delta_{1}}}\right)\right]
\end{align}

In the $m\to\infty$ case the order parameters $q_{0}$ and $\hat{q}_{0}$
need to be rescaled with $m$ and reparametrized with two new quantities
$\delta q_{0}$ and $\delta\hat{q}_{0}$, as follows:
\begin{align}
q_{0} & =q_{1}-\frac{\delta q_{0}}{m}\\
\hat{q}_{0} & =\hat{q}_{1}-\frac{\delta\hat{q}_{0}}{m}
\end{align}
As a consequence, we also reparametrize $\Delta_{0}$ with a new parameter
$\delta\Delta_{0}$ defined as:
\begin{equation}
m\left(\Delta_{1}-\Delta_{0}\right)=\delta\Delta_{0}=\begin{cases}
\frac{2\delta q_{0}}{\pi\sqrt{1-q_{1}^{2}}} & \left(\textrm{sign}\right)\\
\frac{\delta q_{0}}{\pi}\mathrm{arctan\left(\sqrt{\frac{1+q_{1}}{1-q_{1}}}\right)} & \left(\textrm{ReLU}\right)
\end{cases}\label{eq:deltaDelta0}
\end{equation}
The expressions eqs.~(\ref{eq:Gs_sph}), (\ref{eq:Gs_bin}) and~(\ref{eq:Ge})
become:

\section{Distribution of stabilities\label{sec:Distribution-of-stabilities}}

The stability for a given pattern/label pair $\xi^{*},\sigma^{*}$
is defined as:
\begin{align}
\Xi\left(\xi^{*},\sigma^{*}\right) & =\frac{\sigma^{*}}{\sqrt{K}}\sum_{l=1}^{K}c_{l}\,g\left(\sqrt{\frac{K}{N}}\sum_{i=1}^{N/K}W_{li}\xi_{li}^{*}\right)
\end{align}
The distribution over the training set for a typical solutions can
thus be computed as
\begin{equation}
P\left(\Xi\right)=\ensuremath{\left\langle \!\frac{\int\!\mathrm{d}\mu\left(W\right)\,\mathbb{X}_{\xi,\sigma}\left(W\right)\delta\left(\Xi-\Xi\left(\xi^{1},\sigma^{1}\right)\right)}{\int\!\mathrm{d}\mu\left(W\right)\,\mathbb{X}_{\xi,\sigma}\left(W\right)}\!\right\rangle _{\!\xi,\sigma}}
\end{equation}
where we arbitrarily chose the first pattern/label pair $\xi^{1}$,
$\sigma^{1}$ without loss of generality. The expression can be computed
by the replica method as usual, and the order parameters are simply
obtained from the solutions of the saddle point equations for the
free entropy. The resulting expression at the RS level is:
\begin{equation}
P\left(\Xi\right)=\Theta\left(\Xi\right)\ensuremath{\int Dz\,\frac{G\left(\frac{\Xi-z\sqrt{\Delta-\Delta_{-1}}}{\sqrt{\Delta_{2}-\Delta}}\right)}{\sqrt{\Delta_{2}-\Delta}}H\left(-z\sqrt{\frac{\Delta-\Delta_{-1}}{\Delta_{2}-\Delta}}\right)^{-1}}
\end{equation}
where $G\left(x\right)=\frac{1}{\sqrt{2\pi}}e^{-x^{2}/2}$ is a standard
Gaussian. The difference between the models (spherical/binary and
sign/ReLU) is encoded in the different values for the overlaps and
in the different expressions for the parameters $\Delta$, $\Delta_{-1}$,
$\Delta_{2}$.

In the large deviation case, we simply compute the expression with
a $1$RSB ansatz and fix $q_{1}$ and $m$ as described in the previous
section. The resulting expression is
\begin{equation}
P\left(\Xi\right)=\frac{\Theta\left(\Xi\right)}{\sqrt{\Delta_{2}-\Delta_{1}}}\int Dz_{0}\frac{\int Dz_{1}G\left(\frac{\Xi-z_{0}\sqrt{\Delta_{0}-\Delta_{-1}}+z_{1}\sqrt{\Delta_{1}-\Delta_{0}}}{\sqrt{\Delta_{2}-\Delta_{1}}}\right)H\left(-\frac{z_{0}\sqrt{\Delta_{0}-\Delta_{-1}}+z_{1}\sqrt{\Delta_{1}-\Delta_{0}}}{\sqrt{\Delta_{2}-\Delta_{1}}}\right)^{m-1}}{\int Dz_{1}H\left(-\frac{z_{0}\sqrt{\Delta_{0}-\Delta_{-1}}+z_{1}\sqrt{\Delta_{1}-\Delta_{0}}}{\sqrt{\Delta_{2}-\Delta_{1}}}\right)^{m}}
\end{equation}
where the effective parameters $\Delta_{-1}$ , $\Delta_{0}$, $\Delta_{1}$
and $\Delta_{2}$ are the same defined in section~\ref{subsec:1RSB-ansatz}.
In the $m\to\infty$ limit the previous expression reduces to

\begin{equation}
P\left(\Xi\right)=\Theta\left(\Xi\right)\int Dz_{0}\frac{G\left(\frac{\Xi-z_{0}\sqrt{\Delta_{1}-\Delta_{-1}}+z_{1}^{*}\sqrt{\delta\Delta_{0}}}{\sqrt{\Delta_{2}-\Delta_{1}}}\right)}{\sqrt{\Delta_{2}-\Delta_{1}}}H\left(-\frac{z_{0}\sqrt{\Delta_{1}-\Delta_{-1}}+z_{1}^{*}\sqrt{\delta\Delta_{0}}}{\sqrt{\Delta_{2}-\Delta_{1}}}\right)^{-1}
\end{equation}
where $z_{1}^{*}$ satifies

\begin{equation}
z_{1}^{*}=\mathrm{\mathrm{argmax_{z_{1}}\left[-\frac{z_{1}^{2}}{2}+\ln\left(-\frac{z_{0}\sqrt{\Delta_{1}-\Delta_{-1}}+z_{1}\sqrt{\delta\Delta_{0}}}{\sqrt{\Delta_{2}-\Delta_{1}}}\right)\right]}}
\end{equation}
where $\delta\Delta_{0}$ is defined in equation~(\ref{eq:deltaDelta0}).
\end{document}